\documentclass[twocolumn,showpacs,showkeys,amsmath,amssymb,nofootinbib,superscriptaddress,floatfix]{revtex4}

\usepackage{graphicx}
\usepackage{subfigure}
\usepackage{bm}

\makeatletter
\newcommand\erfc{\mathop{\operator@font erfc}\nolimits}
\def\slashchar#1{\setbox0=\hbox{$#1$}
   \dimen0=\wd0 \setbox1=\hbox{/} \dimen1=\wd1
   \ifdim\dimen0>\dimen1 \rlap{\hbox to \dimen0{\hfil/\hfil}} #1
   \else  \rlap{\hbox to \dimen1{\hfil$#1$\hfil}} / \fi}

\newcommand{\lrd}{\raisebox{0.09em}{$\stackrel{\scriptstyle\leftharpoonup\hspace{-.7em}\rightharpoonup}{D} $}}
\newcommand{\ld}{\raisebox{0.09em}{$\stackrel{\scriptstyle\leftharpoonup}{D}$}}
\newcommand{\rd}{\raisebox{0.09em}{$\stackrel{\scriptstyle\rightharpoonup}{D}$}}

\newcommand{\ket}[1]{\left| #1 \right>}
\newcommand{\bra}[1]{\left< #1 \right|}

\newcommand{\I}{\mathrm{i}}

\newcommand{\eVdist}{\kern-0.06667em}
\makeatother

\begin{document}
 
\title{Generalized vector form factors of the pion in a chiral quark model\footnote{Dedicated to the memory of Manoj K. Banerjee. To appear 
in a special issue of the Indian Journal of Physics.}}

\author{Wojciech Broniowski} 
\email{Wojciech.Broniowski@ifj.edu.pl}
\affiliation{The H. Niewodnicza\'nski Institute of Nuclear Physics, Polish Academy of Sciences, PL-31342 Krak\'ow, Poland} 
\affiliation{Institute of Physics, Jan Kochanowski University, %ul.~\'Swi\c{e}tokrzyska 15,
PL-25406~Kielce, Poland}

\date{31 May 2008}

\begin{abstract}
Generalized vector form factors of the pion, related to the moments of the generalized parton distribution functions, are evaluated in the Nambu--Jona-Lasinio model with the 
Pauli-Villars regularization. The lowest moments (the electromagnetic and the gravitational form factors)
are compared to recent lattice data, with fair agreement. Predictions for higher-order moments are also made.  
Relevant features of the generalized form factors in the chiral quark models are highlighted and the role of the QCD evolution 
for the higher-order GFFs is stressed.
\end{abstract}

\pacs{12.38.Lg, 11.30, 12.38.-t}

\keywords{generalized parton distributions, generalized form factors, structure of the pion, chiral quark models, Nambu--Jona-Lasinio model}

\maketitle 

\section{Introduction}

Generalized Parton Distributions (GPDs) are very interesting physical objects, as they 
encode in a natural way the rich information on the internal structure of hadrons (for
extensive reviews see
e.g.~\cite{Ji:1998pc,Radyushkin:2000uy,Goeke:2001tz,
Diehl:2003ny,Ji:2004gf,Belitsky:2005qn,Feldmann:2007zz,Boffi:2007yc}
and references therein). Moments of the GPDs in the $X$ variable form polynomials in the $\xi$ variable (see the following), with the 
$t$-dependent coefficients known as the 
generalized form factors (GFFs). Experimentally, the GPDs, especially for the pion, are rather 
elusive quantities, as they show up in difficult to measure hard exclusive processes such as Deeply Virtual Compton Scattering
(DVCS) or hard electroproduction of mesons (HMP). On the other hand, the GFFs are accessible to lattice QCD measurements. 
In particular, the lowest-order  vector \cite{lat} and tensor \cite{lat2} GFFs of the pion have recently been determined from 
full-QCD lattice calculations 
and more similar results are announced to appear in the future.  

In this paper we evaluate the vector GFFs of the pion in a chiral quark model, namely the Nambu--Jona-Lasinio (NJL) model with the 
Pauli-Villars (PV) regularization, using the techniques described in Ref.~\cite{BAG}. Importantly, the results for the GPDs obtained in this 
model obey all formal requirements, such as normalization, proper support, polynomiality and positivity constraints. At the same time, the 
obtained expressions have a non-trivial form which is not of a factorizable in the $t$-variable. 
We compare the lowest form factors, namely the electromagnetic and the gravitational form factor,
to the recent lattice data \cite{lat} and find good qualitative agreement. Predictions for the higher-order form factors, which will be
be confronted to data when more lattice results appear, are also shown. A careful analysis of the role of the QCD evolution 
is presented, which is essential for higher-order GFFs.

\section{Generalized parton distributions and generalized form factors}

\begin{figure}[tb]
\subfigure{\includegraphics[width=5.8cm]{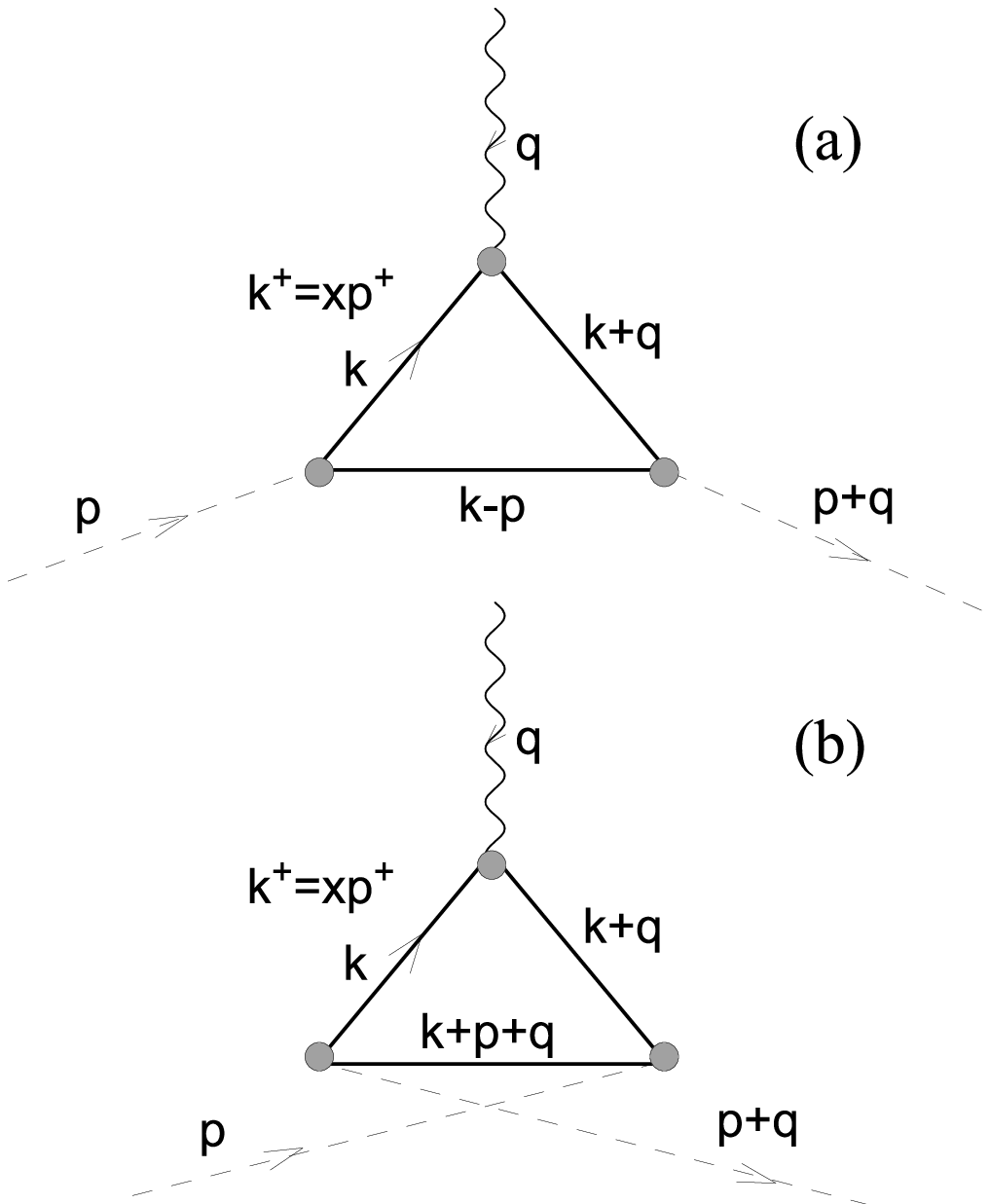}}\\
\subfigure{\includegraphics[width=3.7cm]{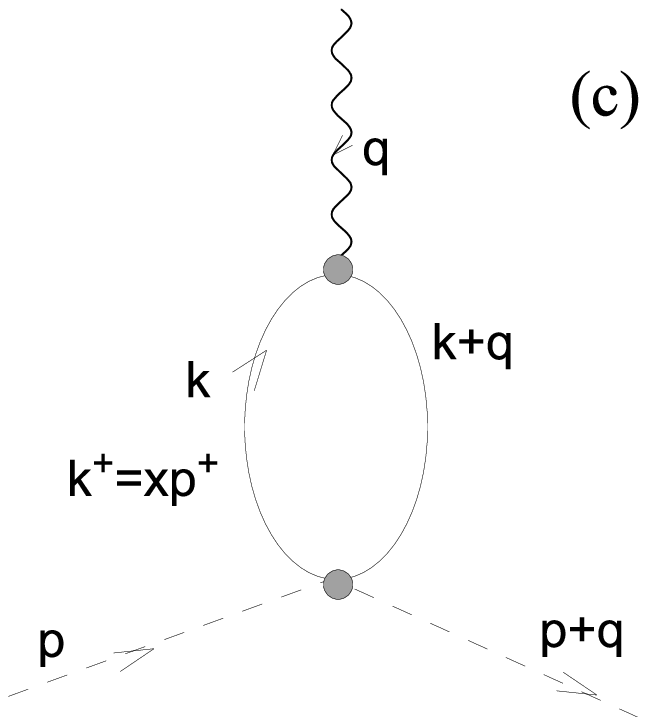}} 
\vspace{-2mm}
\caption{The direct (a), crossed (b), and contact (c) Feynman diagrams for the
quark-model evaluation of the GPD of the pion. The contact contribution is responsible for the $D$-term.}
\label{fig:diag} 
\end{figure}

We begin with the description of kinematics of GPDs. 
The assignment of the momenta in the considered process is depicted in Fig.~\ref{fig:diag}. The standard notation is 
\begin{eqnarray}
&& p^2=m_\pi^2, \;\; q^2=-2p\cdot q=t,\nonumber \\ && n^2=0, \;\; p
\cdot n=1, \;\; q \cdot n=-\zeta. \label{kin}
\end{eqnarray}
The null vector $n$ defines the light cone.
Throughout this paper we work for simplicity in the strict chiral limit with 
\begin{eqnarray}
m_\pi=0,
\end{eqnarray}
 although an extension to the 
physical pion mass is straightforward in the NJL model. 
The two isospin projections of the GPDs of the pion are defined through the matrix elements 
of bilinears of quark fields displaced along the light cone, namely
\begin{eqnarray}
&& \delta_{ab}\,{\cal H}^{I=0}(x,\zeta,t) = \int \frac{dz^-}{4\pi}
e^{i x p^+ z^-} \times \\ && \;\;\; \left . \langle \pi^b (p+q) | \bar
\psi (0) \gamma \cdot n \psi (z) | \pi^a (p) \rangle \right
|_{z^+=0,z^\perp=0}, \nonumber \\ && i \epsilon_{3ab}\,{\cal
H}^{I=1}(x,\zeta,t) = \int \frac{dz^-}{4\pi} e^{i x p^+ z^-} \times
\label{defGPD01} \\ && \;\;\; \left . \langle \pi^b (p+q) | \bar \psi
(0) \gamma \cdot n \psi (z) \, \tau_3 | \pi^a (p) \rangle \right
|_{z^+=0,z^\perp=0}, \nonumber
\end{eqnarray} 
where $z$ is on the light cone, and $a$ and $b$ are the pion isospin indices. At the scale pertaining to the chiral 
quark models the gluons are integrated out, hence ${\cal H}^{G}(x,\zeta,t)=0$.

In chiral quark models at the leading-$N_c$ level the calculation of the GPDs proceeds according to the 
one-loop diagrams 
of Fig.~\ref{fig:diag}. Extensive details of the quark-model evaluation are given in \cite{BAG}. 
The pion GPDs were also analyzed in other variants of chiral quark models in 
Refs.~\cite{Praszalowicz:2002ct,Tiburzi:2002kr,Tiburzi:2002tq,Broniowski:2003rp,Praszalowicz:2003pr,Theussl:2002xp}. In this paper
we use the non-linear NJL model with the PV regularization in the twice-subtracted version proposed in
Ref.~\cite{RuizArriola:2002wr}, where an observable ${\cal O}$ is regularized according to the prescription 
\begin{eqnarray}
{\cal O}_{\rm reg} = {\cal O}(0) - {\cal O}(\Lambda^2 ) + \Lambda^2
\frac{d {\cal O}(\Lambda^2 )}{d\Lambda^2}. \label{prescr}
\label{eq:PV2} 
\end{eqnarray} 
The quantity $\Lambda$ is the PV regulator. In what follows we take $M=280$~MeV for the
constituent quark mass and $\Lambda=871$~MeV, which yields $f=93.3$~MeV for the pion decay constant
\cite{RuizArriola:2002wr} according the the formula
\begin{eqnarray}
f^2=-\frac{3M^2}{4\pi^2} \left ( \log(\Lambda^2 + M^2) \right )_{\rm reg}. \label{f2njl}
\end{eqnarray} 
Calculations in other variants of the chiral quark models, in particular in the Spectral Quark Model \cite{RuizArriola:2001rr,RuizArriola:2003bs},
will be presented elsewhere. They are qualitatively similar to the results presented here.

The pion electromagnetic form factor in the NJL model is equal to
\begin{eqnarray}
&&F_V^{\rm NJL}(t)=1+ \frac{N_c M^2}{8 \pi^2 f^2} \times \label{ffnjl} \\
&&\left ( 
\frac{2 \sqrt{4 \left(M^2+\Lambda ^2\right)-t} \log \left(\frac{\sqrt{4 \left(M^2+\Lambda ^2\right)-t}-\sqrt{-t}}{\sqrt{4
   \left(M^2+\Lambda ^2\right)-t}+\sqrt{-t}}\right)}{\sqrt{-t}}
\right )_{\rm reg}, \nonumber
\end{eqnarray}
The condition $\lim_{t \to - \infty} F_{\rm NJL}(t)=0$ is satisfied due to
Eq.~(\ref{f2njl}).  

In the so-called {\em symmetric} notation for the GPDs, more convenient for our study, one introduces
\begin{eqnarray}
\xi= \frac{\zeta}{2 - \zeta}, \;\;\;\; X = \frac{x - \zeta/2}{1 -
\zeta/2}, \label{xiX}
\end{eqnarray}
where $0 \le \xi \le 1$ and $-1 \le X \le 1$. Then one defines
\begin{eqnarray}
H^{I=0,1}(X,\xi,t)={\cal H}^{I=0,1}\left ( \frac{\xi + X}{\xi + 1},
\frac{2 \xi}{\xi + 1},t \right ),
\end{eqnarray}
which exhibit the reflection properties about the $X=0$ point,
\begin{eqnarray}
H^{I=0}(X,\xi,t)&=&-H^{I=0}(-X,\xi,t), \nonumber \\
H^{I=1}(X,\xi,t)&=&H^{I=1}(-X,\xi,t). \label{eq:sym}
\end{eqnarray} 
In addition, for $X \ge 0$ one has
\begin{eqnarray}
{H}^{I=0,1}(X,0,0) = q(X), \nonumber
\label{eq:q} 
\end{eqnarray} 
relating the distributions to the the pion's forward diagonal parton
distribution function (PDF), $q(X)$.

The {\em polynomiality} conditions~\cite{Ji:1998pc,Radyushkin:2000uy}
state that the moments of the GPDs can be written as
\begin{eqnarray}
\int_{-1}^1 \!\!\!\!\! dX\,X^{2j} \, {H}^{I=1}(X,\xi,t) = 2\sum_{i=0}^j A_{2j+1,i}(t) \xi^{2i}, \nonumber \\
\int_{-1}^1 \!\!\!\!\! dX\,X^{2j+1} \, {H}^{I=0}(X,\xi,t) = 2\sum_{i=0}^{j+1} A_{2j+2,i}(t) \xi^{2i}, \label{poly}
\end{eqnarray}
where $A_{2j+1,i}(t)$ are the generalized form factors, depending on $j=0,1,\dots$ and $i$.  The polynomiality property
follows from very basic field-theoretic assumptions such as the Lorentz
invariance, time reversal, and hermiticity, hence is automatically
satisfied in approaches that obey these requirements. In our approach
polynomiality is manifest from the use of the double distributions \cite{BAG}.

The notation and normalization factors in Eq.~(\ref{poly}) are adjusted in order to agree with 
the conventions of Ref.~\cite{lat}, except for the subscript $i$ which in our case labels the powers of $\xi^2$ and not $\xi$.
We note that for the isovector (non-singlet) GPD only even, and for the isoscalar (singlet) GPD only odd moments are non-zero. 
For the few lowest values of $j$ we have explicitly
\begin{eqnarray}
&&\int_{0}^1 \!\!\!\!\! dX\,        {H}^{I=1}(X,\xi,t) = A_{10}(t)=F_V(t), \label{norm} \\
&&\int_{0}^1 \!\!\!\!\! dX\, X   \, {H}^{I=0}(X,\xi,t) = A_{20}(t)+A_{21}(t)\xi^2 \nonumber \\
&& \hspace{3.5cm}= \frac{1}{2}\theta_2(t)-\frac{1}{2} \theta_1(t) \xi^2, \label{norm2} \\
&&\int_{0}^1 \!\!\!\!\! dX\, X^2 {H}^{I=1}(X,\xi,t) = A_{30}(t)+A_{31}(t)\xi^2, \nonumber\\
&&\int_{0}^1 \!\!\!\!\! dX\, X^3 {H}^{I=0}(X,\xi,t) = A_{40}(t)+A_{41}(t)\xi^2 +A_{42}(t)\xi^4.\nonumber
\end{eqnarray}
In Eq.~(\ref{norm}) we have introduced is the electromagnetic form factor $F_V(t)\equiv A_{10}(t)$, while $\theta_1(t)=2A_{20}$
and $\theta_2(t)=-2A_{21}$ in Eq.~(\ref{norm2}) are the gravitational form factors of the pion, discussed in more detail 
in \cite{BAG}. In the chiral limit these form factors satisfy the low energy theorem
$\theta_1(0) =\theta_2(0)$ \cite{Donoghue:1991qv}. In our quark model calculation in the chiral limit
\begin{eqnarray}
\theta_1(t) =\theta_2(t) \equiv \theta(t),
\end{eqnarray} 
hence, consequently,
$A_{20}(t)=-A_{21}(t)$. 
The sum rule (\ref{norm}) expresses the
electric charge conservation, while (\ref{norm2}) is responsible for
the momentum sum rule in the deep inelastic scattering. 

\begin{figure}[tb]
\includegraphics[width=.48\textwidth]{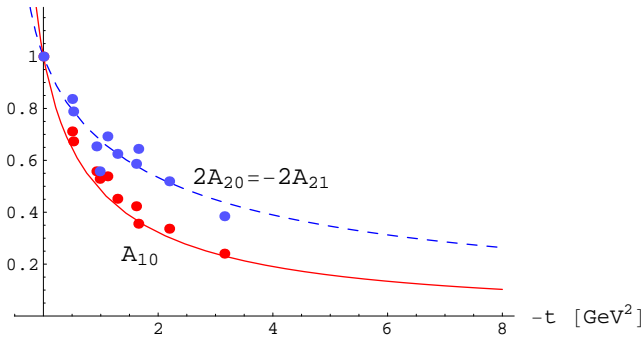} 
\vspace{-2mm}
\caption{The vector and gravitational form factors of the pion in the NJL model with PV
regularization in the chiral limit (lines) compared to the lattice data 
from Ref.~\cite{lat} (statistical errors of the data not displayed). \label{fig:gv}}
\end{figure}

Formulas (\ref{poly}) are equivalent to the definition
\begin{eqnarray}
 \label{eq:gff}
 && \bra{\pi^{+}(p')} \overline{u}(0)\, \gamma^{\{\mu}\, \I\lrd\/^{\mu_1}
  \I\lrd\/^{\mu_2} \dots \I\lrd\/^{\mu_j\}} \,u(0) \ket{\pi^{+}(p)} = \nonumber \\
 && \;\;\;\;2 P^{\{\mu}P^{\mu_1} \dots P^{\mu_j\}} A_{j+1,0}(t)
    +  \\ && \;\;\;\;2\sum^j_{\substack{i=1\\\textrm{odd}}} q^{\{\mu}
    q^{\mu_1} \dots q^{\mu_i}
    P^{\mu_{i+1}} \dots P^{\mu_j\}} \,A_{j+1,(i+1)/2}(t), \nonumber
\end{eqnarray}
where $P=(p+p')/2$,  $\lrd = \frac{1}{2} (\rd -\ld)$, and $\{\dots\}$ denotes symmetrization and subtraction of
traces for each pair of indices. Equivalence of Eq.~(\ref{eq:gff}) and (\ref{poly}) is easily seen by contracting (\ref{eq:gff}) with the null 
vectors $n^{\mu_1} \dots n^{\mu_j}$ and applying the definitions (\ref{kin}).

\section{Results at the quark-model scale}

The calculation of the GPDs made in the NJL model according to the diagrams of Fig.~\ref{fig:diag} 
is straightforward \cite{BAG} and is most efficiently done via the double distributions. Then the the GFFs are extracted from 
the full GPDs by evaluating the moments (\ref{poly}). 
The obtained expressions are rather lengthy, hence we do not present them here. They have the form similar 
to Eq.~(\ref{ffnjl}), involving logs and rational functions in the $t$ variable. 

Our determination of GPDs and GFFs corresponds to 
the {\em quark model scale} $Q_0$, where matching to QCD is made. The reader is referred to Ref.~\cite{BAG},
where the issue is discussed in detail. The 
value of $Q_0$ may be estimated be performing the evolution of the parton distribution functions (PDF) or the parton distribution
amplitude (PDA) to higher scales and comparing the results to the available data. It turns out to be low, $Q_0 \simeq 320$~MeV. 

The results for the electromagnetic and gravitational form factors are independent of the scale. They are shown in Fig.~\ref{fig:gv}. The dots 
are the data points from lattice calculations of Ref.~\cite{lat}.  One 
has to bare in mind that the full-QCD lattice calculations need to be extrapolated to the chiral limit. Also, our model 
incorporates only the leading-$N_c$ contributions. 
Nevertheless, we note a rather remarkable qualitative agreement, 
in particular the feature of a  much slower decay of the gravitational 
form factor compared to the electromagnetic form factor occurs both in
the model and the data. Of course, for the electromagnetic form factor accurate experimental data could could be used for the comparison.
However, the focus of this work is on higher-order GFFs, where the information comes from the lattices. 

\begin{figure}[tb]
\subfigure{\includegraphics[width=.48 \textwidth]{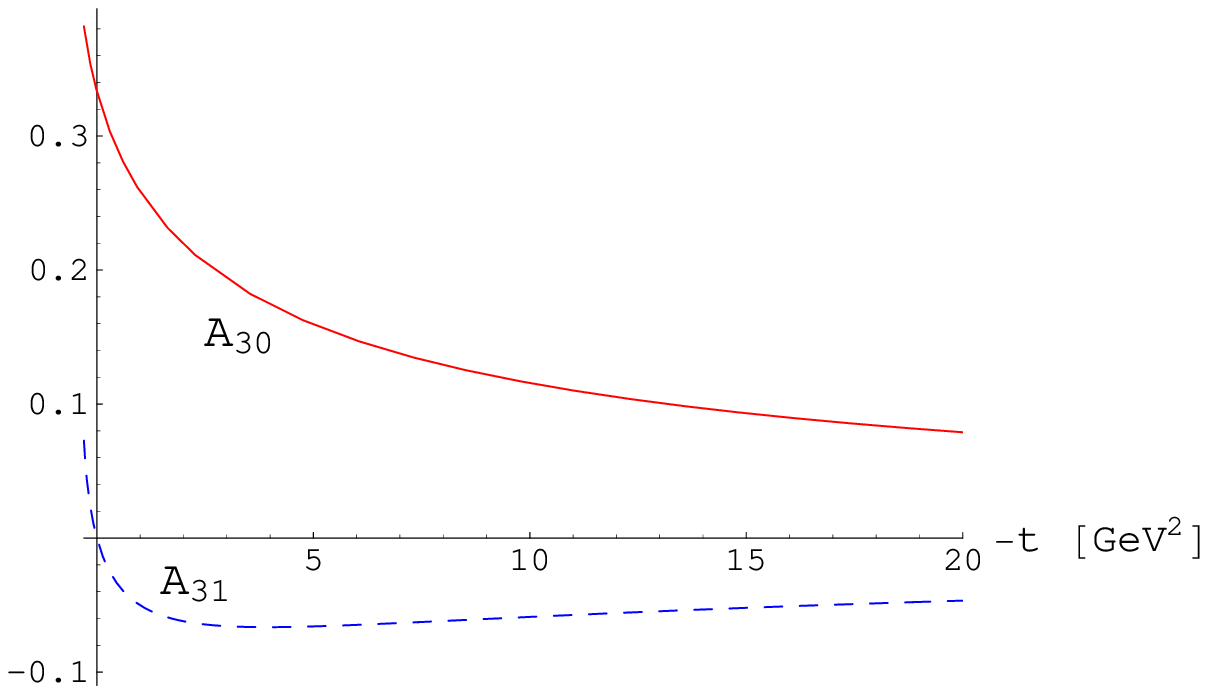}}\\ \vspace{-19mm}
\subfigure{\includegraphics[width=.48 \textwidth]{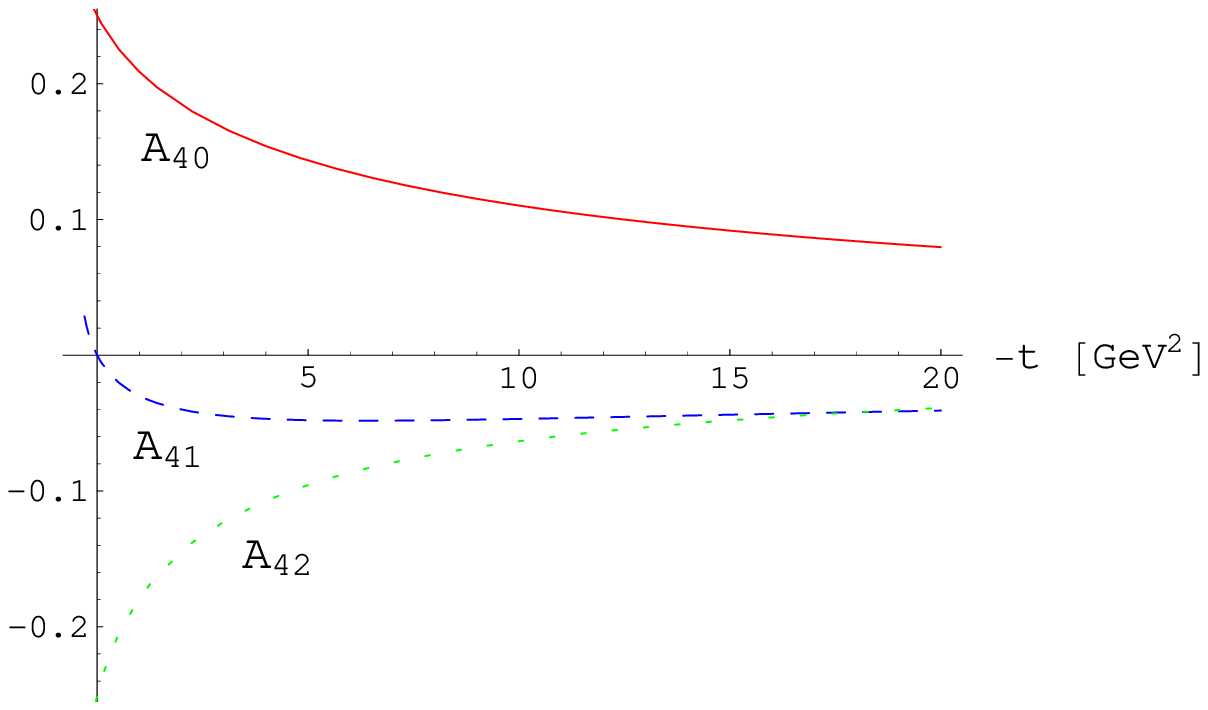}}\\
\vspace{-7mm}
\caption{Generalized vector form factors $A_{3,i}$ and $A_{4,i}$ of the pion in the NJL model with PV regularization at the quark-model scale 
$Q_0$ in the chiral limit. \label{fig:mom34}}
\end{figure}

In Fig.~\ref{fig:mom34} we show the higher level GFFs, $A_{3,i}$ and $A_{4,i}$ obtained at the quark model scale. 
Since in chiral quark models in the chiral limit one has at $t=0$ \cite{Polyakov:1999gs,Theussl:2002xp,BAG} 
\begin{eqnarray}
&& {H}^{I=1}(X,\xi,0) = \theta \left(1-X^2\right)  \\
&& {H}^{I=0}(X,\xi,0) =  \nonumber \\
&& \;\;\;\; \theta ((1-X) (X-\xi))-\theta ((X+1) (-\xi-X)),\nonumber
\end{eqnarray}
it follows from the definition (\ref{poly}) that at the quark-model scale
\begin{eqnarray}
A_{2j+1,i}(0)&=& \left \{ \begin{array}{cl} \frac{1}{2j+1} & {\rm for}\; i=0 \\ 0 & {\rm otherwise}\end{array}\right . \nonumber \\
A_{2j+2,i}(0)&=& \left \{ \begin{array}{cl} \frac{1}{2j+2} & {\rm for}\; i=0 \\ -\frac{1}{2j+2} & {\rm for}\; i=j+1 \\ 
0 & {\rm otherwise}\end{array}\right . \nonumber \\
&& \hspace{1.3cm}{\rm (quark-model~scale)}
\end{eqnarray}
This behavior is seen in Fig.~\ref{fig:mom34}. 
We note that the model GFFs go to zero very slowly at large $-t$. 
Another property follows from the fact that in the considered model ${H}^{I=0}(X,1,t)=0$ for any value of $t$. Then Eq.~(\ref{poly})
yields  
\begin{eqnarray}
\sum_{i=0}^{j+1} A_{2j+2,i}(t)=0.
\end{eqnarray}
This feature can be seen in the lower panel of Fig.~\ref{fig:mom34}.

\section{QCD evolution\label{sec:qcd}}

\begin{figure}[tb]
\subfigure{\includegraphics[width=.48 \textwidth]{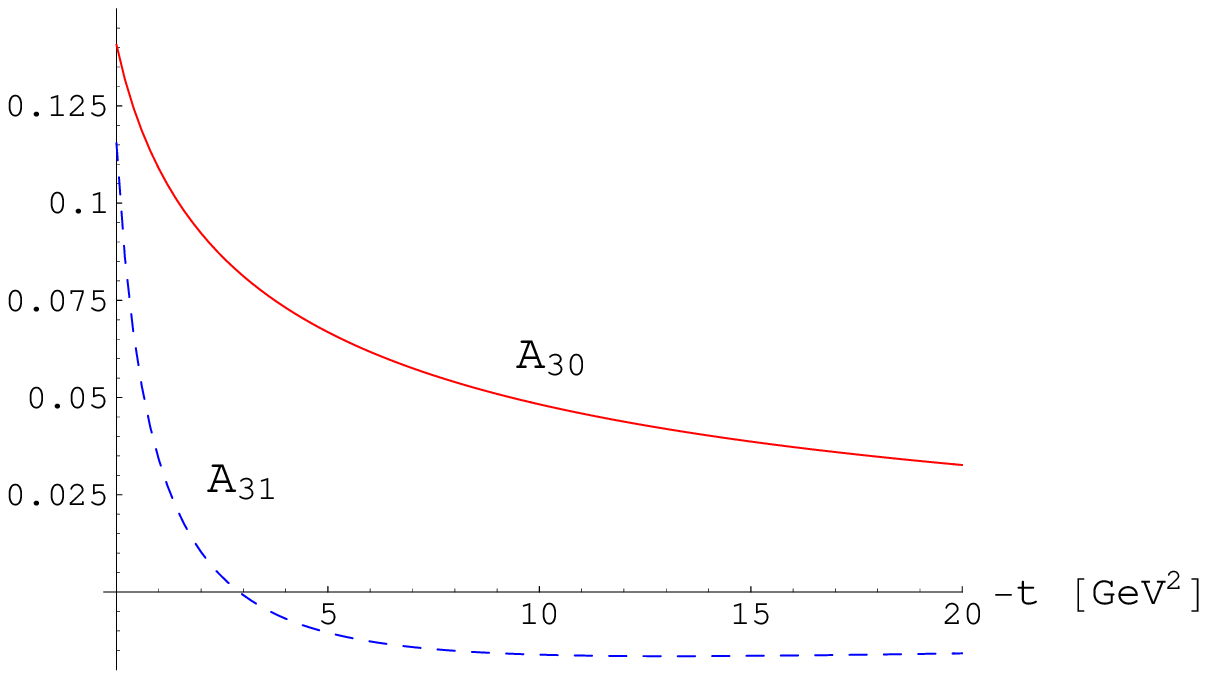}}\\ \vspace{-19mm}
\subfigure{\includegraphics[width=.48 \textwidth]{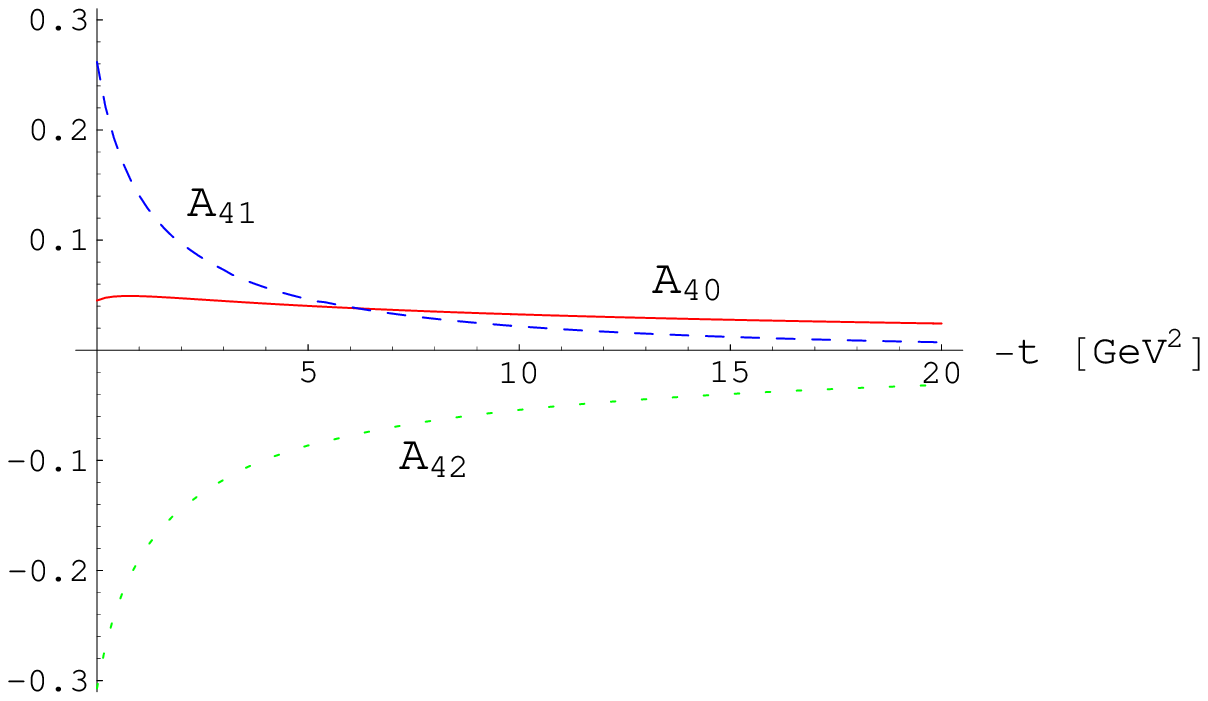}}\\
\vspace{-7mm}
\caption{Same as Fig.~\ref{fig:mom34}
evolved to the lattice scale of Ref.~\cite{lat},
$Q^2=0.71$~GeV$^2$. \label{fig:mom34e}}
\end{figure}

As already mentioned, the crucial role of the QCD evolution in chiral quark model calculations has been 
discussed in Ref.~\cite{BAG}. We carry the leading-order DGLAP-ERBL evolution from the quark-model scale 
\begin{eqnarray}
Q_0=313~{\rm MeV}
\end{eqnarray} 
to the scale of the lattice calculation of Ref.~\cite{lat}. This scale can be inferred from the value 
$2A_{20}(t)=0.63$ in \cite{lat}, which is reproduced in our calculation when we evolve 
the isoscalar GPD to the scale $Q^2=0.71$~GeV$^2$.
Thus we estimate the lattice scale as 
\begin{eqnarray}
Q=843~{\rm MeV}.
\end{eqnarray}
We use the method and code described in \cite{GolecBiernat:1998ja} to evolve the GPDs taken at several selected values of $\xi$. 
From this one may disentangle the coefficients of the powers of $\xi$, namely the GFFs, at the lattice scale $Q$. 
The results of this procedure are presented in Fig.~\ref{fig:mom34e}.
We note a sizable change, both in the value at $t=0$ and in shape, 
compared to the behavior of Fig.~\ref{fig:mom34}, which shows that in general the GFFs do 
evolve with the scale, except for the protected form factors as those in Fig.~\ref{fig:gv}, which are invariants of the evolution.

For the evolution scale $Q^2\to \infty$, the GPDs tend to their asymptotic forms
located entirely in the ERBL region $|X|<\xi$. Explicitly, we have in this limit\cite{BAG}
\begin{eqnarray}
\label{eq:asy}
&&H^{I=1}= \frac{3}{2\xi} \left(1-\frac{X^2}{\xi^2}\right) F_V(t) 
\\
&&H^{I=0}= (1 - \xi^2) \frac{15}{4 \xi^2} \frac{N_f}{4 C_F + N_f} \frac{X}{\xi}\left(1-\frac{X^2}{\xi^2}\right)\theta(t) \nonumber
\\
&&X H^G= (1 - \xi^2)  \frac{15}{4 \xi} \frac{C_F}{4C_F + N_f} \left(1-\frac{X^2}{\xi^2}\right)^2 \theta(t),\nonumber
\end{eqnarray} 
where $C_F=(N_c^2-1)/(2N_c)$ and $N_f=3$ is the number of active flavors.
The proportionality factors reflect the normalization
at the initial quark-model scale $Q_0$, as the  charge- and momentum-conservation sum rules
are invariants of the evolution,
\begin{eqnarray}
&&\int_{-1}^{1}dX\,H^{I=1}(X,\xi,t,Q^2)=2F_V(t),\\
&&\int_{-1}^{1}dX\,\left( X H^{I=0}(X,\xi,t,Q^2)+X H_g(X,\xi,t,Q^2)\right) \nonumber \\
&&=(1-\xi^2) \theta(t), \nonumber
\end{eqnarray}
in accordance to Eq.~(\ref{norm},\ref{norm2}).

Evaluation of moments in Eq.~(\ref{eq:asy}) yields for $Q^2\to \infty$
\begin{eqnarray}
A_{2j+1,i}(t)&=& \left \{ \begin{array}{cl} \frac{3}{4j(j+2)+3} F_V(t) & {\rm for}\; i=j 
                                                      \\ 0 & {\rm otherwise}\end{array}\right . \nonumber \\
A_{2j+2,i}(t)&=& \left \{ \begin{array}{cl} \frac{N_f}{4 C_F + N_f}\frac{15}{2[4j(j+4)+15]} \theta(t) & {\rm for}\; i=j 
                                                      \\ -A_{2j+2,j}(t) & {\rm for}\; i=j+1 \\
                                                                      0 & {\rm otherwise}\end{array}\right . \nonumber \\
A^G_{2j+2,i}(t)&=& \frac{4C_f}{N_f} \frac{1}{2j+1} A_{2j+2,i}(t), \nonumber \\
&& \hspace{21mm} (Q^2\to \infty) \label{asGFF}                                                              
\end{eqnarray}
with the gluon form factors defined as 
\begin{eqnarray}
\int_{-1}^{1}dX\,X^{2j+1} H_g(X,\xi,t,Q^2)=2\sum_{i=0}^{j+1} A^G_{2j+2,i}(t)\,\xi^{2j}.
\end{eqnarray}
We note a striking difference of the asymptotic form factors compared to the form factors at the quark model scale shown in Fig.~\ref{fig:mom34}.
For the isovector case only the highest form factor, for $i=j$, is non-zero, and for the isoscalar case only the two highest moments, for 
$i=j+1$ and $i=j$ are non-zero. In contract, at the quark-model scale all GFFs are present. Thus only $A_{10}$ and $A_{20}=-A_{21}$ are 
invariants of the QCD evolution.
The asymptotic gluon form factors in Eq.~(\ref{asGFF}) are related to the isoscalar quark form factors in a simple manner. 
Asymptotically, all GFFs become
proportional to $F_V(t)$ or $\theta(t)$ in the isovector and isoscalar channels, respectively. The universality for the gluon form factors follows from the fact that we start from the quark-model initial condition, which carries no gluons, which are then built in the process 
of evolution. Asymptotically, for $N_c=3$ and $N_f=3$ 
the quark to gluon ratio in the momentum sum rule equals $9/16$.  

\section{Summary}

In summary, we have computed the generalized vector form factors of the pion in the Nambu--Jona-Lasinio model with the Pauli-Villars
regularization. We have proceeded through the $X$-moments of the generalized parton distribution functions, evolved from the quark-model scale to 
the scale of the lattice calculations. The model GPDs exhibit  no factorization in the $t$-variable. 
Comparison to the lattice results for the electromagnetic and gravitational form factors of the the pion have 
been made, with proper agreement. The QCD evolution has been carried out and its role for the higher-order generalized form factors has been 
discussed. Our predictions for higher-order GFFs may be compared to lattice results when these become available.  

\section{Manoj}

Since this volume is devoted to the memory of Manoj K. Banerjee, our unforgettable teacher and friend, let me finish with an anecdote.
When I started my graduate work, Manoj gave me a code written together with Mike Birse who was then a postdoc at Maryland. The code was used to
solve the chiral soliton model with valence quarks (the Birse-Banerjee model \cite{BirBan,BirBan2}), which was a major achievement. My assignment was to introduce vector mesons to the model \cite{BB,BB2}.
When reading down the FORTRAN lines I noticed that the encoded radial differential equation for the pion had a seriously-looking mistake: in one of the terms 
instead of {\tt p/r**2}
there was {\tt p/r*2}. Omission of one asterisk changed the square into the multiplication by 2, a potentially devastating error. When I showed 
this to my advisor, he rushed to the computer and ran the corrected code. The first output quantity was the soliton mass, which changed by a tiny amount at the relative level of $10^{-4}$ or so. 
Other observables were also practically unaffected \dots -- ``You see, young man, good physics is immune to such silly 
mistakes as omission of an asterisk!''

Let Manoj's kindness, brilliance, enthusiasm, and confidence be with us!

\bigskip

\begin{acknowledgments}
This work is a direct extension of the research made with Enrique Ruiz Arriola and Krzysztof Golec-Biernat \cite{BAG}.
The author is grateful to Lech Szymanowski for useful discussions and especially to Krzysztof Golec-Biernat 
for making available the code solving the QCD evolution for the GPDs.
This work has been supported by the Polish Ministry of Science and Higher
Education, grant N202~034~32/0918.
\end{acknowledgments}

%\bibliography{manoj}

\end{document}